\documentstyle[12pt]{article}
\begin{document}
\baselineskip 18pt
\def\today{\ifcase\month\or
 January\or February\or March\or April\or May\or June\or
 July\or August\or September\or October\or November\or December\fi
 \space\number\day, \number\year}
\def\thebibliography#1{\section*{References\markboth
 {References}{References}}\list
 {[\arabic{enumi}]}{\settowidth\labelwidth{[#1]}
 \leftmargin\labelwidth
 \advance\leftmargin\labelsep
 \usecounter{enumi}}
 \def\newblock{\hskip .11em plus .33em minus .07em}
 \sloppy
 \sfcode`\.=1000\relax}
\let\endthebibliography=\endlist
\def\lsim{\ ^<\llap{$_\sim$}\ }
\def\gsim{\ ^>\llap{$_\sim$}\ }
\def\r2{\sqrt 2}
\def\rmuu{\gamma^{\mu}}
\def\rmud{\gamma_{\mu}}
\def\PL{{1-\gamma_5\over 2}}
\def\PR{{1+\gamma_5\over 2}}
\def\sinW2{\sin^2\theta_W}
\def\AEM{\alpha_{EM}}
\def\v#1{v_#1}
\def\tb{\tan\beta}
\def\epem{$e^+e^-$}
\def\KK{$K^0$-$\bar{K^0}$}
\def\wi{\omega_i}
\def\xj{\chi_j}
\def\Wmu{W_\mu}
\def\Wnu{W_\nu}
\def\m#1{{\tilde m}_#1}
\def\mH{m_H}
\def\mw#1{{\tilde m}_{\omega #1}}
\def\mx#1{{\tilde m}_{\chi #1}}
\def\mwi{{\tilde m}_{\omega i}}
\def\mxj{{\tilde m}_{\chi j}}
\def\rwi{r_{\omega i}}
\def\rxj{r_{\chi j}}
\def\rfp{r_f'}
\def\Kij{K_{ij}}
\def\Fq{F(q^2)}
\begin{titlepage}

\  \
\vskip 0.5 true cm 
\begin{center}
{\large {\bf Neutron Electric Dipole Moment }}  \\
{\large {\bf from Supersymmetric Anomalous $W$-boson Coupling }}
\vskip 0.5 true cm
\vspace{2cm}
\renewcommand{\thefootnote}
{\fnsymbol{footnote}}
Tomoko Kadoyoshi 
\footnote{The Doctoral Research Course in Human Culture.}
and Noriyuki Oshimo
\\
\vskip 0.5 true cm 
\it Department of Physics, Ochanomizu University  \\
\it Otsuka 2-1-1, Bunkyo-ku, Tokyo 112, Japan  \\
\end{center}

\vskip 4.0 true cm

\centerline{\bf Abstract}
\medskip
      In the supersymmetric standard model (SSM) 
the $W$-boson could have a 
non-vanishing electric dipole moment (EDM) through a one-loop 
diagram mediated by the charginos and neutralinos.  
Then the $W$-boson EDM induces 
the EDMs of the neutron and the electron.  
We discuss these EDMs, taking into consideration the 
constraints from the neutron and electron EDMs at 
one-loop level induced by 
the charginos and squarks or sleptons.  
It is shown that the neutron and the electron  
could respectively have EDMs of order 
of $10^{-26}e$cm and $10^{-27}e$cm,     
solely owing to the $W$-boson EDM.   
Since these EDMs do not depend on the values of SSM parameters
for the squark or slepton sector, they provide less ambiguous 
predictions for CP violation in the SSM.     

\vskip 0.5 true cm
\noindent 
PACS number(s):  11.30.Er, 12.60.Jv, 13.40.Em, 14.70.Fm, 14.80.Ly

\end{titlepage}

\newpage 
\section{Introduction} 

    Several extensions of the Standard Model (SM) contain new sources 
of CP violation as well as the standard Kobayashi-Maskawa (KM) 
mechanism.  
Although the observed CP-violating phenomena 
in the \KK\ system can be well 
described by the SM, the baryon asymmetry of the universe seems to 
imply the existence of a new source of CP violation.  
Any new discovery of CP-violating phenomenon will 
therefore greatly help us to investigate CP violation, which      
could give an important clue to new 
physics beyond the SM.  
One of such observables is the electric dipole moment (EDM) 
of an elementary particle \cite{EDMrev}.  

      In this paper we study the EDMs  
of the neutron and the electron caused by the EDM of the 
$W$-boson within the framework of the 
supersymmetric standard model (SSM).  
If the $W$-boson has a non-vanishing EDM, it can generally 
induce the neutron and electron EDMs through one-loop 
diagrams generated by the SM interactions \cite{1wedm}.  
These processes could actually occur in the SSM \cite{2wedm}, 
since the $W$-boson EDM receives a contribution from a one-loop 
diagram mediated by the charginos and neutralinos shown in 
Fig. \ref{fig1}, 
owing to a new CP-violating phase 
in the gauge-Higgs sector \cite{ellis}.  
On the other hand, the same CP-violating phase 
leads to the EDMs of the neutron and the electron 
at one-loop level \cite{ellis,edm,edmnew}, 
which are potentially large and thus 
put severe constraints on the SSM.  
Taking into account these constraints, we make detailed 
analyses of the $W$-boson EDM and the resultant EDMs of 
the neutron and the electron.  
Although these neutron and electron EDMs arise from  
two-loop diagrams, we show, 
their magnitudes are not much smaller than 
their present experimental upper bounds.  
Since the KM mechanism in the SM does not 
predict such a large magnitude 
neither for the neutron EDM nor the electron EDM,  
possible detection of these EDMs in near-future experiments 
could make the SSM 
a more credible candidate for the new physics.    

     As stated above, the EDMs of the neutron and the 
electron receive contributions at one-loop level 
in the SSM.  One might think that these mechanisms would dominate the 
EDMs, making those two-loop contributions which are 
related to the $W$-boson EDM negligible.  However, 
it is shown that there are sizable regions of SSM parameter 
space where the two-loop contributions become 
comparable with the one-loop contributions.  
Furthermore, the one-loop contributions depend on the SSM parameters 
both for the gauge-Higgs sector and for the squark or slepton sector, 
while the two-loop contributions depend 
on the former alone.
We can thus predict the values of the neutron and electron 
EDMs induced by the $W$-boson EDM with less uncertainty 
in the SSM.  

\begin{figure}
\vspace{10.5cm}
\includegraphics{wedm1.ps}
\caption{The Feynman diagram for the EDM of the $W$-boson.}
\label{fig1}
\end{figure}         
     This paper is organized as follows:  
In Sec. 2 we will summarize the origin of 
CP violation in the SSM and present the interactions relevant  
to our discussions.  
In Sec. 3 we will discuss an anomalous CP-odd coupling for 
the $W$-bosons and photon generated by the interactions of 
the charginos, neutralinos, and $W$-bosons, 
which leads to the EDM of the $W$-boson.  
In Sec. 4 the EDMs 
of the neutron and the electron will be considered.  
The two-loop contributions caused by the $W$-boson EDM will be 
calculated and analyzed paying attention to the one-loop contributions 
mediated by the charginos and squarks or sleptons.    
Conclusion will be given in Sec. 5.   

\section{Model} 

     The SSM is an extension of the SM based on  
$N=1$ supergravity coupled to grand unified 
theories (GUTs) \cite{SUSYrev}.  
This model contains several parameters 
whose values are generally complex.    
Assuming minimal particle contents, these complex parameters 
are the followings:  
Yukawa coupling constants $\eta_f$; the vacuum expectation values 
$\v1$ and $\v2$ of the 
Higgs bosons with U(1) hypercharges $-1/2$ and $1/2$, respectively;   
the SU(3), SU(2), and U(1) gaugino masses $\m3$, $\m2$, and $\m1$, 
respectively, which appear in supersymmetry soft-breaking terms;  
the mass parameter $\mH$ in a bilinear term of 
Higgs superfields in superpotential;    
and the dimensionless constant $a$, 
which originates in trilinear couplings of the 
squarks or sleptons in supersymmetry soft-breaking terms.   
We may assume that the complex phases of $\m3$, $\m2$, and $\m1$ 
are not different from each other  
because of the coincidence of these values at the GUT scale.  However, 
unless some extra symmetry is imposed on the SSM, 
the other complex phases are independent each other.  
Then, the redefinitions of relevant fields cannot rotate away all the 
complex phases.  
Even if generation-mixings among matter fields 
are neglected, two of the complex parameters cannot be taken real.   
These are the new sources of CP violation intrinsic in the SSM.  

     The EDM of the $W$-boson is induced through a one-loop diagram, 
which is generated by the interactions of 
the charginos, neutralinos, and $W$-bosons.  
The charginos $\wi$ and the neutralinos $\xj$  
are charged and neutral mass eigenstates 
for the gauginos and Higgsinos, whose mass matrices  
are respectively given by 
\begin{equation}
    M^- = \left(\matrix{\m2 & -g\v1^*/\r2 \cr
                -g\v2^*/\r2 & \mH}        \right) 
\label{1} 
\end{equation}
and 
\begin{equation}
    M^0 = \left(\matrix{\m1 &  0  & g'\v1^*/2 & -g'\v2^*/2 \cr
                         0  & \m2 & -g\v1^*/2 &   g\v2^*/2 \cr
                       g'\v1^*/2 & -g\v1^*/2 &   0  & -\mH \cr
                      -g'\v2^*/2 &  g\v2^*/2 & -\mH &   0}
           \right).
\label{2}
\end{equation}
These mass matrices are diagonalized by unitary 
matrices $C_R$, $C_L$, and $N$ as
\begin{equation}
      C_R^\dagger M^-C_L = {\rm diag}(\mw1, \mw2) \quad 
                       (\mw1 <\mw2 )   
\label{3}
\end{equation}
and
\begin{equation}
N^tM^0N = {\rm diag}(\mx1, \mx2, \mx3, \mx4) \quad
                       (\mx1<\mx2<\mx3<\mx4), 
\label{4}
\end{equation}
giving the mass eigenstates.  

     The complex mass matrices for the charginos and 
the neutralinos lead 
to CP-violating interactions of the charginos, neutralinos, 
and $W$-bosons.  
The interaction Lagrangian for these particles is given by 
\begin{eqnarray}
{\cal L}&=&\frac{1}{\r2}g\bar{\xj}\rmuu
                  \left(G_{Lji}\PL+G_{Rji}\PR\right)
                           \wi\Wmu^\dagger +{\rm H.c.},    
\label{5}    \\
& & G_{Lji} = \r2 N^*_{2j}C_{L1i}+N^*_{3j}C_{L2i},  \nonumber  \\
 & & G_{Rji} = \r2 N_{2j}C_{R1i}-N_{4j}C_{R2i}.  \nonumber  
\end{eqnarray}   
Since the coupling constants $G_{Lji}$ and $G_{Rji}$ have 
different 
complex phases, this Lagrangian is not invariant under 
CP transformation.  
The SSM parameters which determine the 
interactions in eq. (\ref{5}) are $\v1$,
$\v2$, $\m2$, $\m1$, and $\mH$
appearing in eqs. (\ref{1}) and (\ref{2}).  We assume the GUT relation 
$\m1=(5/3)\tan^2\theta_W\m2$.  The redefinitions of 
the fields make it possible without loss of 
generality to take all these parameters 
except $\mH$ real and positive.  Then, 
the CP-violating phase is represented by the phase of $\mH$, 
which we express as  
\begin{equation}
\mH=|\mH|\exp(i\theta).  
\label{6}
\end{equation}
Since $\v1$ and $\v2$ are related to the $W$-boson mass $M_W$, 
independent parameters become $\tb$, $\m2$, $|\mH|$, 
and $\theta$, where the ratio of the vacuum expectation values 
is denoted by $\tb$ ($\equiv \v2/\v1$).  

     In the SSM the EDM of the neutron  
receives contributions from one-loop diagrams in which 
the charginos, neutralinos, or gluinos are exchanged together 
with the squarks.  
The electron EDM is also induced by one-loop diagrams, 
where the charginos or neutralinos are exchanged with the sleptons.  
Among these diagrams, the chargino-mediated 
ones generally make dominant contributions \cite{edm}.  
The interaction Lagrangian for the charginos, $u$-type quarks, and 
$d$-type squarks and that for the charginos, $d$-type quarks, and 
$u$-type squarks are respectively given by 
\begin{eqnarray}
{\cal L}&=&ig\bar{\wi^c}
                  \left(A_{Li}^k\PL+A_{Ri}^k\PR\right)
                           u{\tilde d}_k^\dagger    +{\rm H.c.},    
\label{7}    \\
& & A_{Li}^k = C_{L1i}S_{d1k}^*
                      -\frac{\eta_d}{g}C_{L2i}S_{d2k}^*,  \nonumber \\
& & A_{Ri}^k = -\frac{\eta_u^*}{g}C_{R2i}S_{d1k}^*     \nonumber  
\end{eqnarray}
and
\begin{eqnarray}
{\cal L}&=&ig\bar{\wi}
                  \left(B_{Li}^k\PL+B_{Ri}^k\PR\right)
                           d{\tilde u}_k^\dagger  +{\rm H.c},
\label{8}   \\
& & B_{Li}^k = C_{R1i}^*S_{u1k}^*
                     +\frac{\eta_u}{g}C_{R2i}^*S_{u2k}^*,  \nonumber \\ 
& & B_{Ri}^k = \frac{\eta_d^*}{g}C_{L2i}^*S_{u1k}^*.     \nonumber  
\end{eqnarray}
Here $S_f$ and $\eta_f$ $(f=u, d)$ represent 
respectively the unitary matrix 
which diagonalizes the 
mass-squared matrix for the $f$-type squarks and 
the Yukawa coupling constant 
for the $f$-type quark.  
The generation-mixings have been neglected.  
The matrices $S_u$ and $S_d$ depend on various SSM parameters, though 
they can be approximated by the unit matrix 
for the $u$- and $d$-squarks  
because of the smallness of the corresponding quark masses.  
Then, the SSM parameters which determine 
the interactions in eqs. (\ref{7}) and (\ref{8}) are the squark masses 
and those which determine the interactions in eq. (\ref{5}).   
The interaction Lagrangians for the charginos, 
leptons, and sleptons are 
obtained by appropriately changing eqs. (\ref{7}) and (\ref{8}), 
for which additional parameters are principally the slepton masses.   

         The EDMs of the neutron and the electron 
induced at one-loop level 
could be as large as their present experimental upper bounds 
$|d_n|\lsim 10^{-25}e$cm \cite{nEDM} 
and $|d_e|\lsim 10^{-26}e$cm \cite{eEDM}, respectively.  
Accordingly, the values of the SSM parameters are constrained.  
In order to satisfy the experimental bounds, 
either the CP-violating phases  
are small or the squarks and sleptons are heavy.  
Since there is no known mechanism which naturally suppresses 
the complex phases of the SSM parameters, 
we take that the CP-violating phases are of order unity.    
Under this assumption the masses of the  
squarks and sleptons are found to be larger than 1 TeV, 
whereas those of the charginos and  
neutralinos can be of order of 100 GeV \cite{edm}.  
It is also shown that $\tb$ is unlikely to have a value 
much larger than 10.

\section{EDM of $W$-boson} 

\begin{figure}
\vspace{10.5cm}
\includegraphics{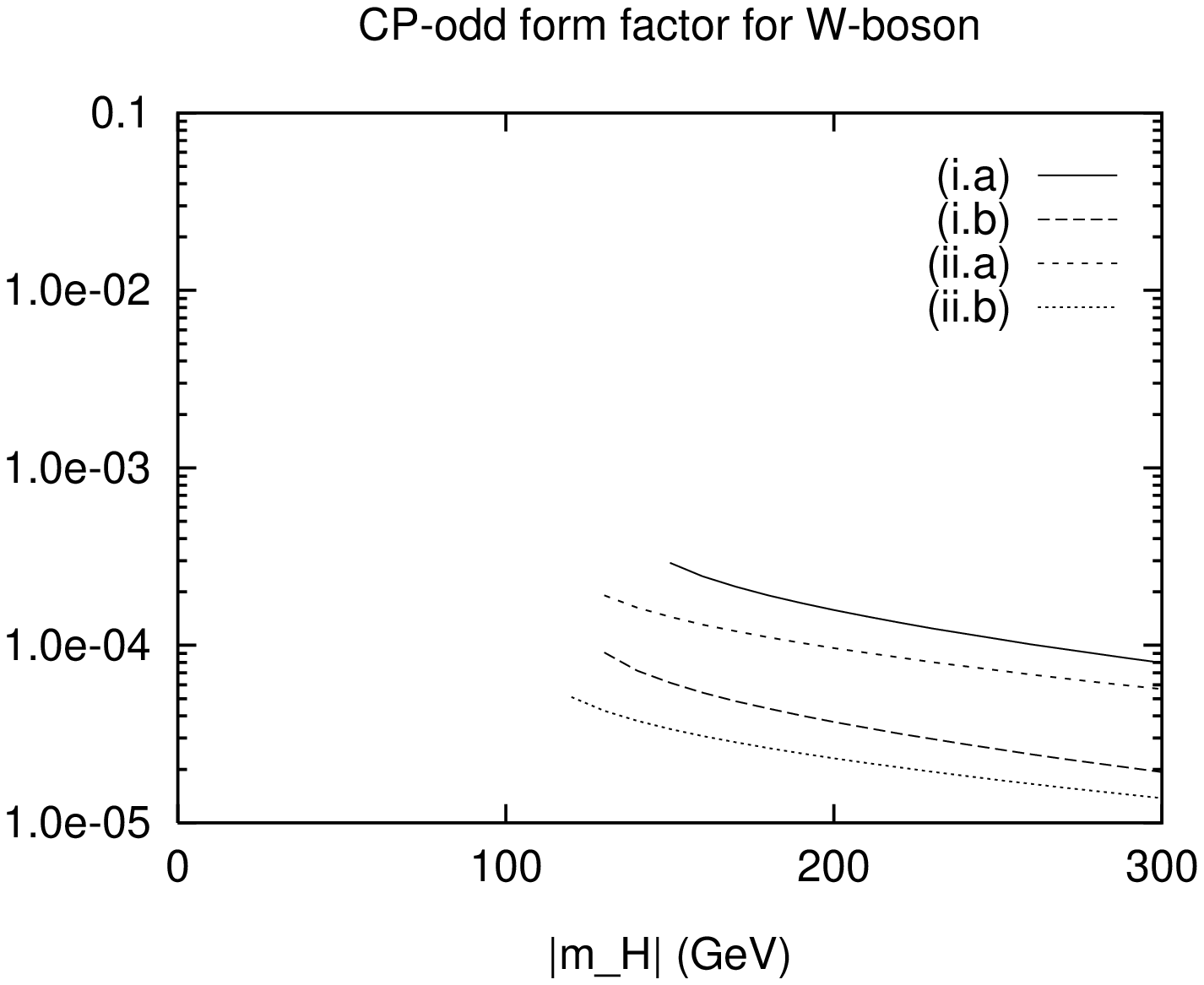}
\caption{The CP-odd form factor of the $WW\gamma$ 
         interactions for $\theta=\pi/4$.  
        The values of $\m2$ and $\tb$ for curves (i.a)--(ii.b) 
         are given in Table 1.}
\label{fig2}
\end{figure}         
    The interactions in eq. (\ref{5}) give rise to 
a CP-violating anomalous coupling of the $W$-bosons and  
photon through the one-loop diagram shown in Fig. \ref{fig1}, 
where the charginos and neutralinos are exchanged.  
The CP-violating term in the effective Lagrangian is given by  
\begin{eqnarray}
& & {\cal L}_{eff}=ie\Fq W_\mu^\dagger W_\nu\frac{1}{2}
\epsilon^{\mu\nu\rho\sigma}
(\partial_\rho A_\sigma -\partial_\sigma A_\rho),        
\label{9} \\
& & \Fq=-\frac{\AEM}{2\pi\sinW2}\sum_{i=1}^{2}\sum_{j=1}^{4}
{\rm Im}(G_{Lji}^*G_{Rji})\frac{\mwi\mxj}{M_W^2} \nonumber \\
& &  \int_{0}^{1}dx\int_{0}^{1-x}dy
[x(1-x)+y(1-x-y)r-x\rxj-(1-x)\rwi]^{-1},
\label{10}
\end{eqnarray}
\[
r=\frac{q^2}{M_W^2}, \quad \rwi=\frac{\tilde m_{\omega i}^2}{M_W^2}, 
\quad \rxj=\frac{\tilde m_{\chi j}^2}{M_W^2}, 
\]
where $q^2$ denotes the momentum-squared of the photon.  The $W$-boson 
has been put on mass-shell.    
For the $W$-bosons and a vector boson, in general,  
there can be two more couplings which break CP invariance \cite{hagi}.  
However, the interactions in eq. (\ref{5}) 
only contribute to the coupling of eq. (\ref{9}).  
The EDM of the $W$-boson is given in terms of 
the CP-odd form factor $\Fq$ by 
\begin{equation}
   d_W=-\frac{e}{2M_W}F(0).
\label{11}
\end{equation}

\begin{table}
\begin{center}
\begin{tabular}{c|c|c|c|c}
\hline
    & (i.a) & (i.b) & (ii.a) &(ii.b) \\
\hline
 $\m2$ (GeV) & 200 & 200 & 300 & 300 \\
\hline
 $\tb$  & 2  & 10  & 2 & 10  \\
\hline
\end{tabular} 
\end{center}
\caption{The values of $\m2$ and $\tb$ for curves 
            (i.a)--(ii.b) in Figs. 2, 4, 6, and 7.} 
\label{tab1}
\end{table}
     In Fig. \ref{fig2} we show the value of $\Fq$ as a function of 
the absolute value of the mass parameter $\mH$ 
defined in eq. (\ref{6}).  
For the CP-violating phase we take $\theta=\pi/4$.  
The four curves correspond to four sets of parameter values 
for $\m2$ and $\tb$ given in Table \ref{tab1}.  
The value of $q^2$ is set for $q^2=($200 GeV$)^2$, 
although $\Fq$ does not vary much with $q^2$.  
In the ranges of $|\mH|$ where curves are not 
drawn, the lighter chargino has a mass smaller than  
$(1/2)\sqrt{q^2}$, making pair production of the charginos 
possible through the virtual photon.  
We can see that the magnitude of $\Fq$ is around $10^{-4}$ for 
$\m2, |\mH|\sim 200$ GeV and $\tb=2$.  
Larger values for $\tb$ suppress $\Fq$. 
Since the SM does not contribute to $\Fq$ at one-loop level, 
its prediction for $\Fq$ is much smaller than $10^{-4}$.   
Therefore, the measurement of $\Fq$ will provide an interesting test 
for the SSM.  
It should be noted that the interactions in eq. (\ref{5}) also induce 
CP-violating couplings for the 
$W$-bosons and $Z$-boson at one-loop level, 
which are of the same order of magnitude as $\Fq$ \cite{cpasy}.  

     The CP-violating $WW\gamma$ 
and $WWZ$ couplings could 
in principle be measured in \epem\ 
colliding experiments \cite{schild}.  
Indeed, 
the experiments at LEPII or planned linear colliders 
will explore some anomalous couplings for the $W$-boson.  
For instance, measurements of angular correlations among the 
final fermions produced from a pair of $W$-bosons may be able to  
disclose CP violation coming from the triple gauge boson vertices.  
The possibility of 
examining the CP-violating couplings 
in \epem\ experiments will be discussed 
in another paper \cite{cpasy}.  

\section{EDMs of neutron and electron}

\begin{figure}
\vspace{10.5cm}
\includegraphics{wedm3.ps}
\caption{The Feynman diagram for the EDM of a quark or a lepton 
              which involves a CP-violating coupling for the 
              $W$-bosons and photon.}
\label{fig3}
\end{figure}         
     The EDM of the $W$-boson yields the EDMs of the quarks and 
the leptons through one-loop diagrams.  In the SSM, therefore, 
the EDMs of the quarks and the leptons 
can be induced at two-loop level.  
The relevant diagram is shown in Fig. \ref{fig3},  
where $f$ and $f'$ denote the quarks or leptons whose  
left-handed components form an SU(2) doublet.   
Since the experiments 
can explore  
fairly small values 
for  
the neutron and the electron EDMs, 
the $W$-boson EDM may be observed through  
these EDMs.  

    The EDM of a quark or a lepton $f$ arising from the 
two-loop diagram in Fig. \ref{fig3} 
is given by  
\begin{eqnarray}
 d_f&=&\mp e\left(\frac{\AEM}{4\pi\sinW2}\right)^2
	 \sum_{i=1}^2\sum_{j=1}^4
{\rm Im}(G_{Lji}^*G_{Rji})\frac{\mwi\mxj}{M_W^2}\frac{m_f}{M_W^2} 
  \nonumber \\
& & \frac{1}{2(1-\rfp)^2}\int_{0}^{1}\frac{ds}{s}
  [\{\frac{3-\rfp}{1-\rfp}\cdot\frac{1}{\Kij-\rfp}
                +\frac{1}{(\Kij-\rfp)^2}\}\rfp^2\log\frac{\rfp}{\Kij} 
  \nonumber \\
& & +\{\frac{1-3\rfp}{1-\rfp}\cdot\frac{1}{\Kij-1}
        +\frac{\rfp}{(\Kij-1)^2}\}\log\frac{1}{\Kij} \nonumber \\
& & +\left(\frac{1}{\Kij-\rfp}+\frac{1}{\Kij-1}\right)\rfp],  
\label{12}
\end{eqnarray}
\[
           \Kij=\frac{\rwi}{s}+\frac{\rxj}{1-s},   
\]
\[
 \rfp=\frac{{m'}_f^2}{M_W^2}, 
 \quad \rwi=\frac{\tilde m_{\omega i}^2}{M_W^2}, 
\quad \rxj=\frac{\tilde m_{\chi j}^2}{M_W^2}.  
\]
Here $m_f$ and ${m'}_f$ represent the masses of $f$ and  
$f'$, respectively.  
The negative and positive signs in front of the right side of 
the equation are respectively for the fermions with 
the third components of their weak isospins $1/2$ and $-1/2$. 
At two-loop level there also exist other diagrams 
which involve squarks or 
sleptons and make contributions to the quark or lepton EDM.  
However,
as long as the squarks and sleptons are much heavier than 
the charginos and neutralinos, these diagrams can be safely 
neglected.  
Assuming the nonrelativistic quark model, 
the EDM of the neutron $d_n$ is given, in terms of 
the $u$-quark EDM $d_u$ and the $d$-quark EDM $d_d$, 
by $d_n=(4d_d-d_u)/3$.  

\begin{figure}
\vspace{10.5cm}
\includegraphics{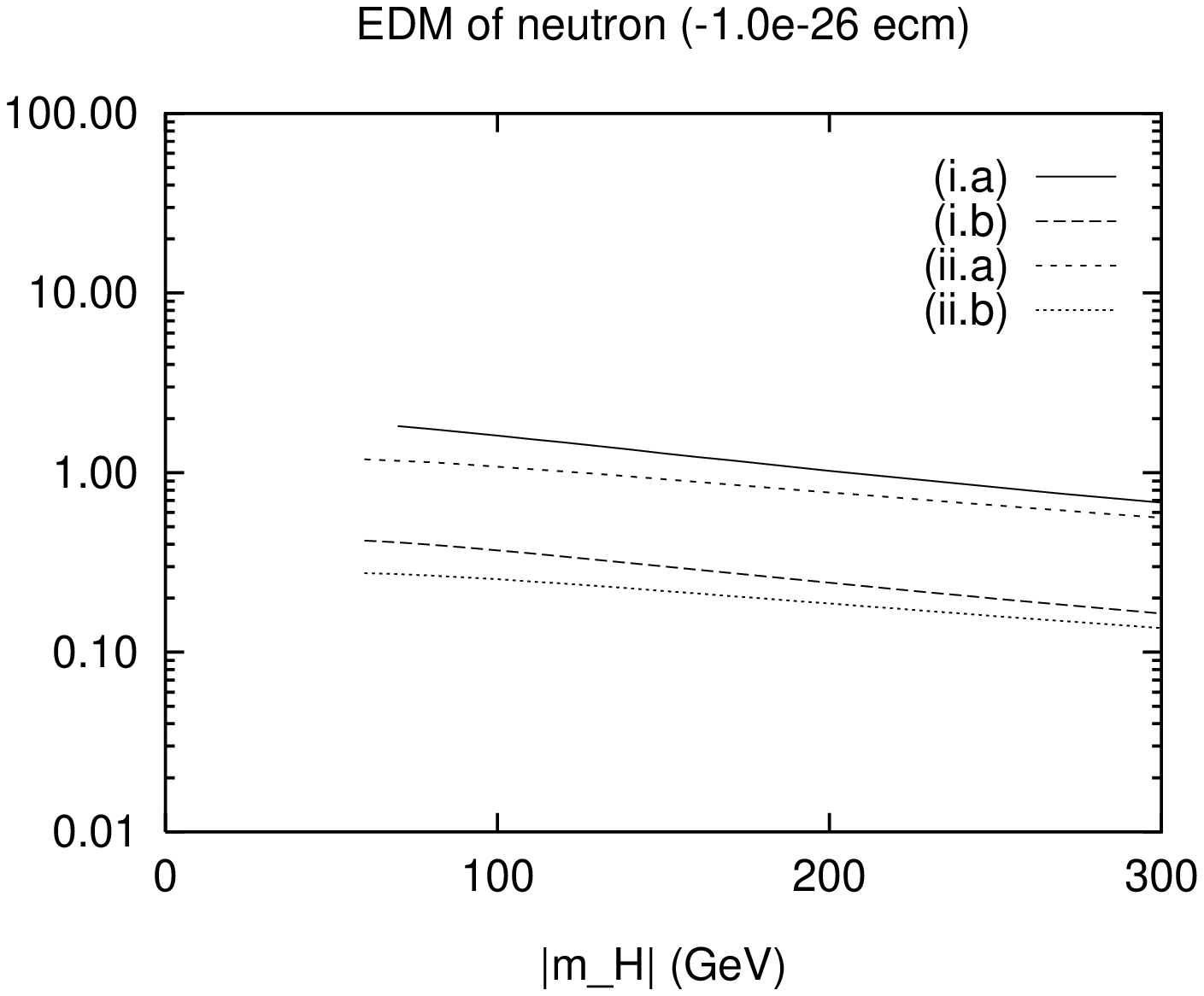}
\caption{The EDM of the neutron induced by the 
        $W$-boson EDM for $\theta=\pi/4$.  
        The values of $\m2$ and $\tb$ for curves (i.a)--(ii.b) 
         are given in Table 1.}
\label{fig4}
\end{figure}
\begin{figure}
\vspace{10.5cm}
\includegraphics{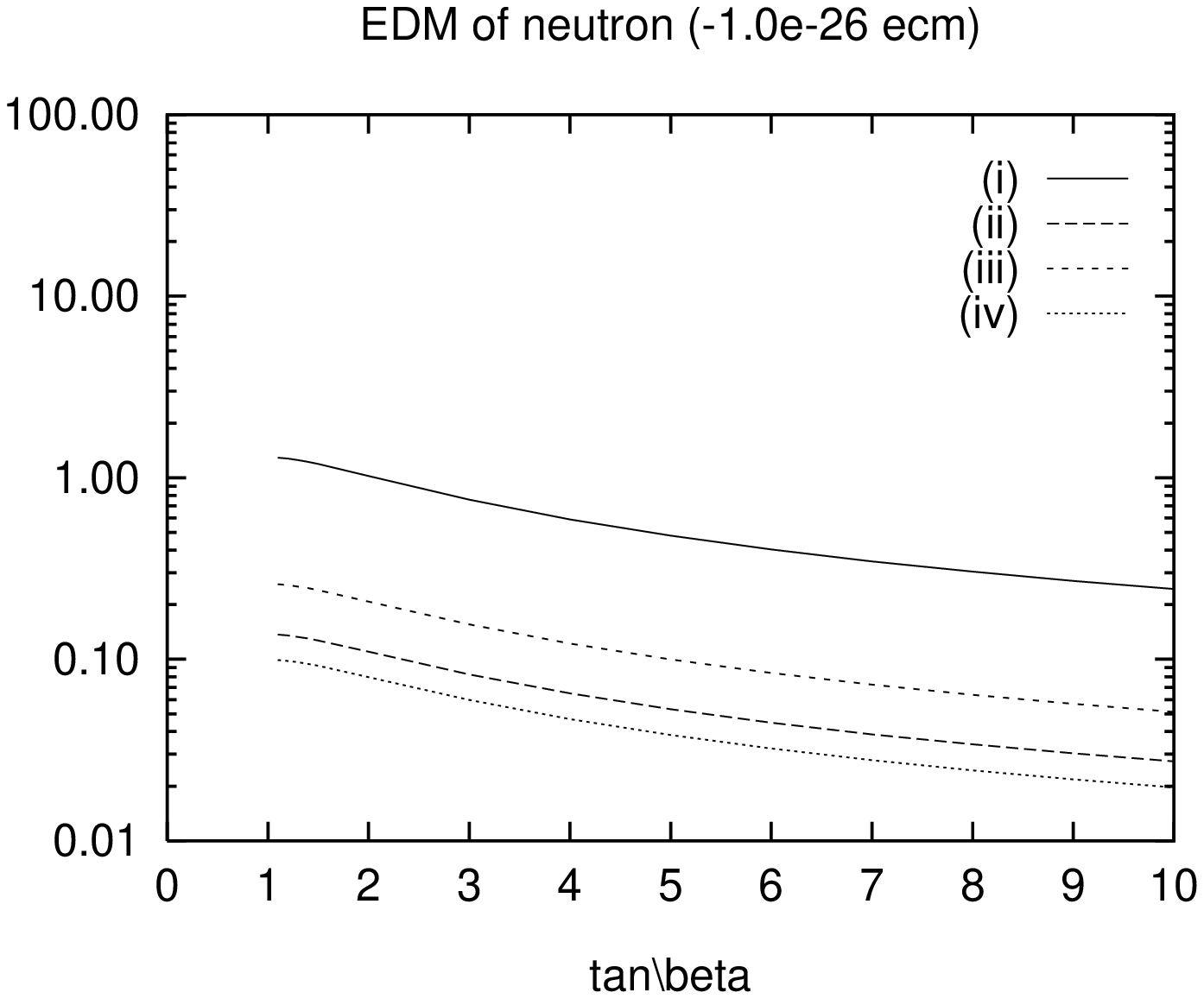}
\caption{The EDM of the neutron induced by the 
        $W$-boson EDM for $\theta=\pi/4$. 
        The values of $\m2$ and $|\mH|$ for curves (i)--(iv) 
         are given in Table 2.}
\label{fig5}
\end{figure}

         
     We show in Fig. \ref{fig4} the numerical value of 
the neutron EDM as a function 
of the absolute value of $\mH$. 
For $\m2$ and $\tb$ we have taken four sets of values given in 
Table \ref{tab1}, corresponding to the four curves in Fig. \ref{fig2}.  
In the ranges of $|\mH|$ where curves are not drawn, 
the lighter chargino has a mass smaller than 45 GeV which has been 
ruled out by LEP experiments.  
In Fig. \ref{fig5} the neutron EDM is plotted as a function of $\tb$ 
for four sets of values of $\m2$ and $|\mH|$ 
given in Table \ref{tab2}.  
Curves are not drawn for $\tb<1$, because 
the value of $\tb$ is theoretically considered not to be smaller than 
unity, if the SU(2)$\times$U(1) gauge symmetry is broken through 
radiative corrections.  
The CP-violating phase is fixed as $\theta=\pi/4$ in 
both Figs. \ref{fig4} and \ref{fig5}.  
For $\m2, |\mH|\sim 200$ GeV and $\tb\sim 2$, 
where the CP-odd form factor for 
the $W$-boson is of order of $10^{-4}$, 
the magnitude of the neutron EDM is around 
$10^{-26} e$cm, which is smaller than the present 
experimental upper bounds by only one order of magnitude.  
The EDM of the neutron decreases as $\m2$ or $|\mH|$ increases.  
We can also see its clear dependence on $\tb$.  
As $\tb$ increases, the EDM decreases.  
This is contrary to the $\tb$-dependence of the 
neutron and electron EDMs induced by one-loop 
diagrams \cite{edm} or the radiative 
$b$-quark decay \cite{bsg} in the SSM.  
Since the squarks should have masses larger than 1 TeV in 
our scheme, 
the magnitude of the neutron EDM arising from two-loop diagrams 
with the squarks becomes much smaller than $10^{-26} e$cm.   

\begin{figure}
\vspace{10.5cm}
\includegraphics{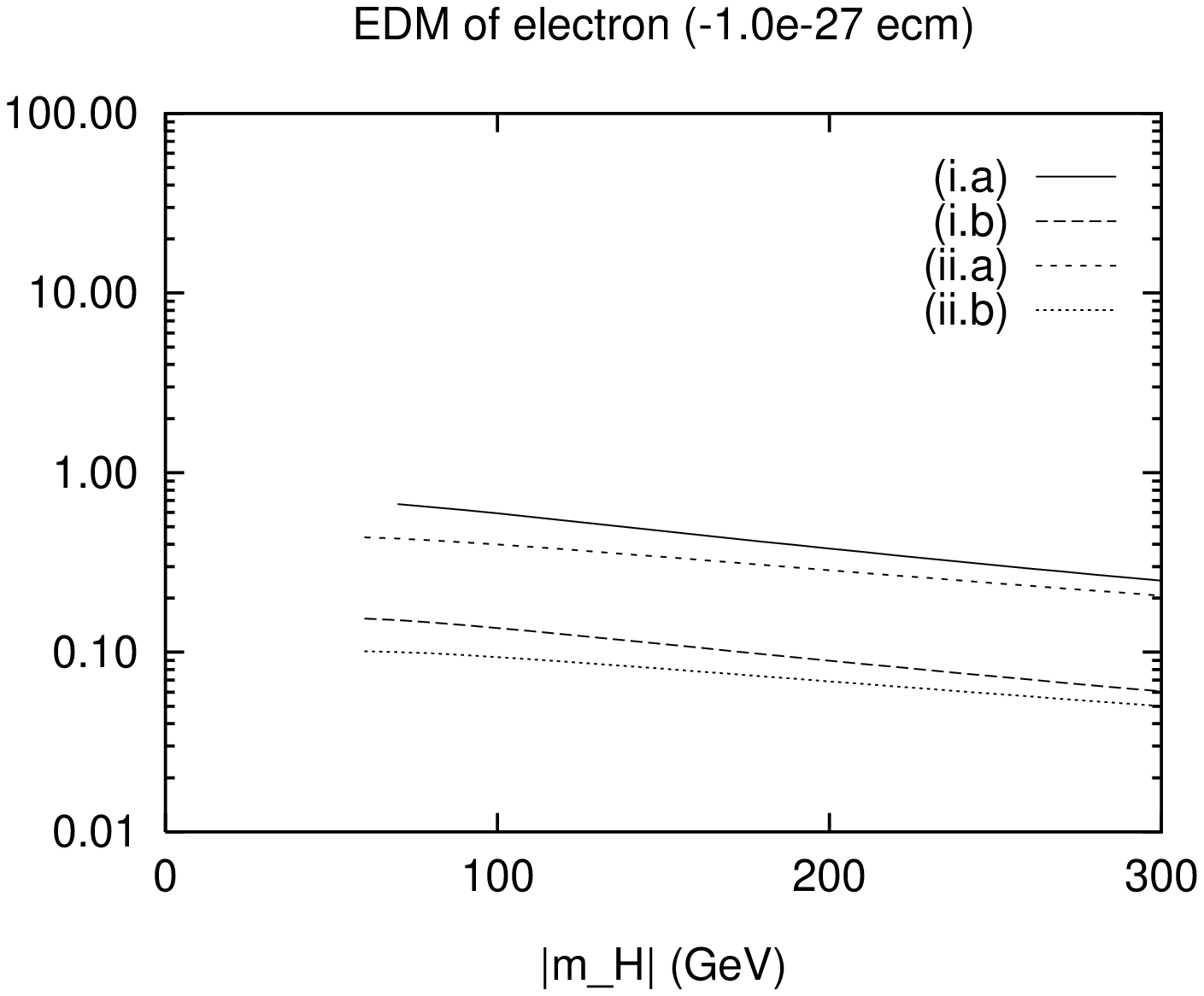}
\caption{The EDM of the electron induced by the 
           $W$-boson EDM for $\theta=\pi/4$.     
         The values of $\m2$ and $\tb$ for curves (i.a)--(ii.b) 
         are given in Table 1.}
\label{fig6}
\end{figure}

\begin{table}
\begin{center}
\begin{tabular}{c|c|c|c|c}
\hline
    & (i) & (ii) & (iii) &(iv) \\
\hline
 $\m2$ (GeV) & 200 & 200 & 1000 & 1000 \\
\hline
 $|\mH|$  (GeV) & 200  & 1000  & 200 & 1000  \\
\hline
\end{tabular} 
\end{center}
\caption{The values of $\m2$ and $|\mH|$ for curves 
                (i)--(iv) in Fig. 5.    }
\label{tab2}
\end{table}
     In Fig. \ref{fig6} we show the value of the electron EDM as a 
function of $|\mH|$ for the values of 
$\m2$ and $\tb$ given in Table \ref{tab1}.  
For $\m2, |\mH|\sim 200$ GeV and $\tb\sim 2$, the electron EDM 
is smaller than its experimental upper bounds 
by one order of magnitude.  
The EDM of the electron is only different from that of the $d$-quark 
by the ratio of their masses.  

     If the neutron and electron EDMs 
respectively have values of order of 
$10^{-26}e$cm and $10^{-27}e$cm, 
they will possibly be detected in the near future.  
In the SM, the EDM of the neutron vanishes 
at both one-loop and two-loop
levels, resulting in $|d_n|<10^{-30}e$cm, 
and the EDM of the electron is much smaller \cite{EDMrev}.    
Therefore, the detection of these EDMs in the foreseeable future   
may be considered to suggest a sizable value for 
the $W$-boson EDM and provide an indirect evidence for 
the existence of supersymmetry in nature.  

\begin{figure}
\vspace{10.5cm}
\includegraphics{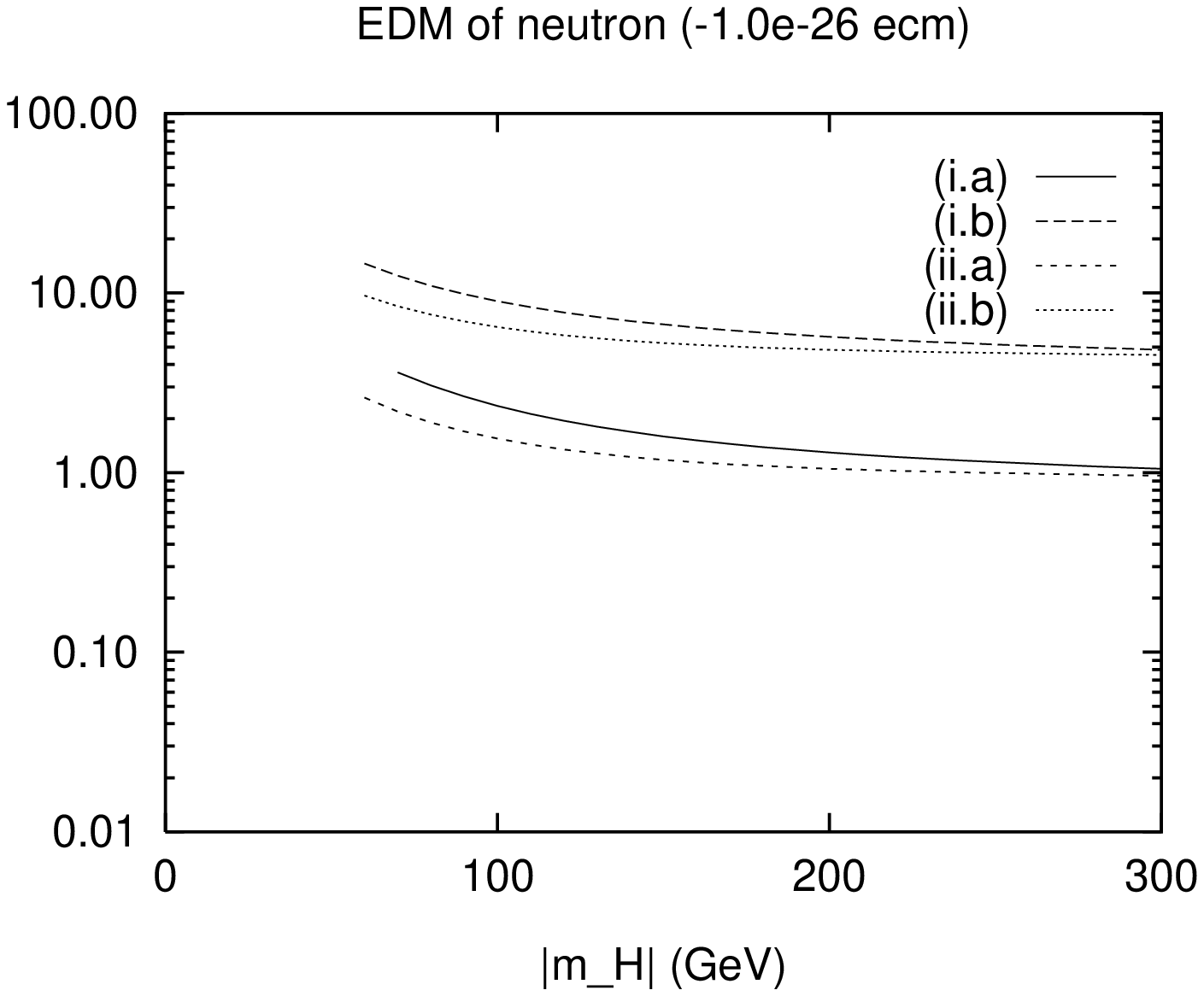}
\caption{The EDM of the neutron from the one-loop diagrams 
         mediated by the charginos and squarks for 
         the squark masses of 10 TeV and $\theta=\pi/4$.    
        The values of $\m2$ and $\tb$ for curves (i.a)--(ii.b) 
         are given in Table 1.}
\label{fig7}
\end{figure}         
     We now make a  
comparison between the above results for the neutron 
and electron EDMs and those at one-loop level in the SSM.  
The EDMs of the neutron and the electron 
receive contributions from the one-loop diagrams mediated by  
the charginos, which generally dominate over other 
one-loop diagrams mediated by the gluinos or neutralinos \cite{edm}.   
These one-loop contributions are a priori expected to be larger 
than the contributions from the two-loop diagrams.  
However, the two-loop contributions 
related to the $W$-boson EDM are determined 
only by the gauge-Higgs sector, while the one-loop contributions 
are determined not only by the gauge-Higgs sector 
but also by the squark or slepton sector, 
especially by their masses.    
Consequently, as the squark or slepton masses  
increase, the two-loop contributions become dominant.  
In Fig. \ref{fig7} we show the neutron EDM 
induced by the chargino-mediated one-loop diagrams for 
the squark masses of 10 TeV and $\theta=\pi/4$.  
Four curves correspond to the 
parameter values for $\m2$ and $\tb$ in Table \ref{tab1}.   
The EDM is found to have a value comparable 
with that from the two-loop contributions 
shown in Fig. \ref{fig4} for $\m2$, $|\mH|\sim 200$ GeV and $\tb=2$.  
In fact, for the squark and slepton masses of around 1 TeV, 
the one-loop contributions to the neutron and electron EDMs 
are larger than the two-loop contributions, whereas for those masses  
of around 10 TeV or larger, the latter  
can become larger than the former.  

\begin{figure}
\vspace{10.5cm}
\includegraphics{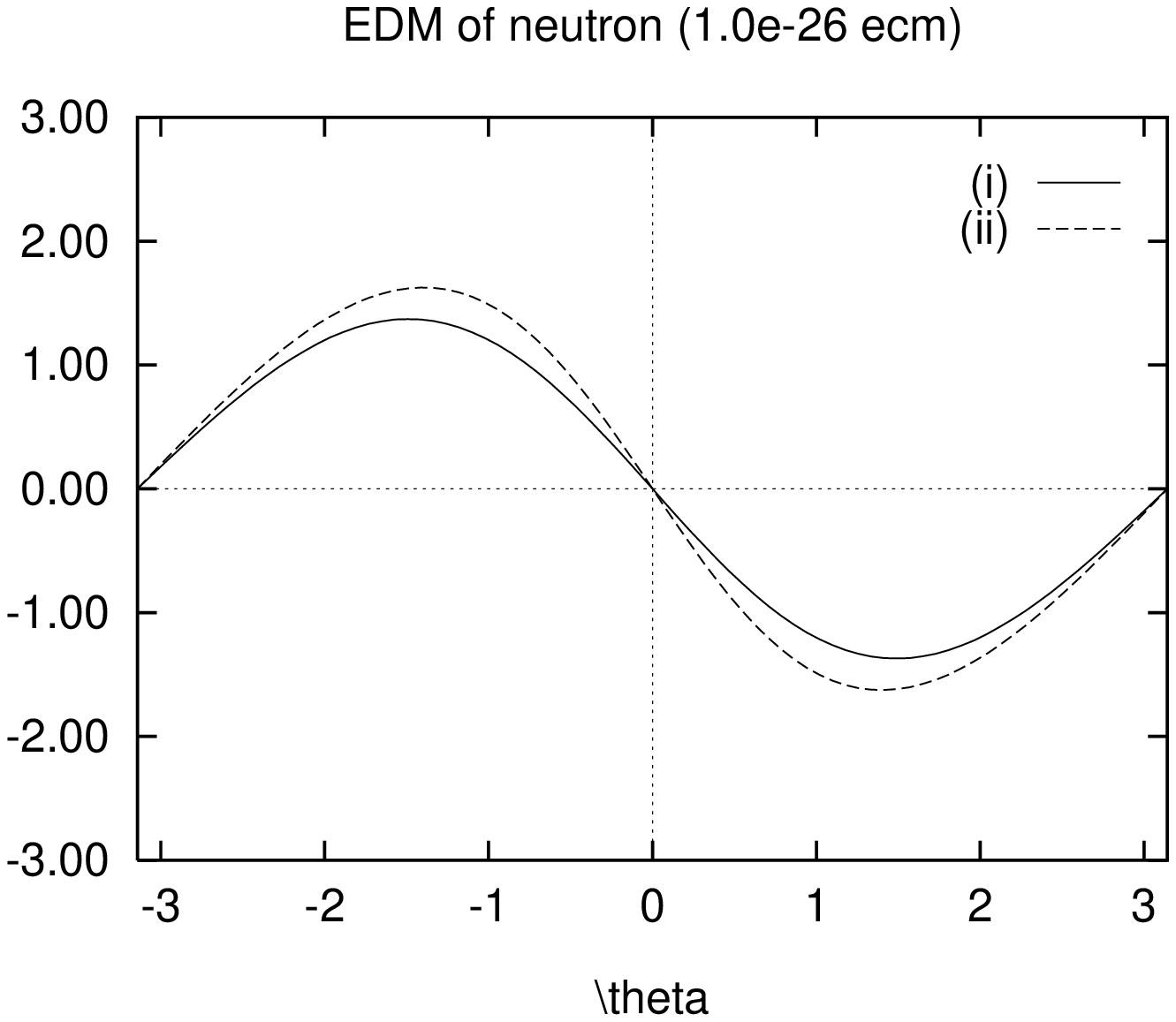}
\caption{The neutron EDM from the two-loop contributions 
      and that from the one-loop contributions,  
      which are represented respectively by curves (i) 
      and (ii).  
        The parameter values are $\m2=|\mH|=200$ 
        GeV and $\tb=2$.  For curve (i) the squark masses 
        are set for 10 TeV.}
\label{fig8}
\end{figure}         
     In Fig. \ref{fig8} we show the values of the neutron EDM 
at one-loop and two-loop levels as 
functions of the CP-violating phase $\theta$ for $\m2=200$ GeV,  
$|\mH|=200$ GeV, and $\tb=2$.  
Curves (i) and (ii) respectively represent the EDM from 
the two-loop contributions 
and that from the one-loop contributions.   
The squark masses for the one-loop 
contributions have been taken for 10 TeV.  It is seen that 
the one-loop and the two-loop contributions have the same sign,  
which also holds for other reasonable values of the parameters 
and for the EDM of the electron.
Hence, for given values of  
the parameters for the gauge-Higgs sector,     
the magnitudes of the neutron and electron EDMs 
are expected to be larger than those obtained from 
the two-loop diagrams of Fig. \ref{fig3}.   

\section{Conclusion}

     We have studied CP violation which 
originates from the gauge-Higgs sector in the SSM.   
The EDM of the $W$-boson is induced at one-loop level through the 
diagram in which the charginos and neutralinos are exchanged.   
It was shown that the CP-odd form factor 
in the effective Lagrangian 
for the $WW\gamma$ interactions could be of order of $10^{-4}$, 
which is far larger than the SM prediction.  
Therefore, the examination  
of the $W$-boson EDM would be an interesting process to search 
for supersymmetry.  
   
     The possibility of detecting the $W$-boson EDM has been discussed 
through the EDMs of the neutron and the electron.    
These EDMs receive contributions from the two-loop diagrams which 
contain the one-loop diagram for the $W$-boson EDM.  
As a result, the large  
value of the $W$-boson EDM implicates large values for the neutron and  
electron EDMs.  We have shown that  
the magnitudes of the neutron and electron EDMs could be  
as large as $10^{-26}e$cm and $10^{-27}e$cm, respectively.    
These numerical outcomes are not so small compared to the experimental 
upper bounds at present, 
and thus may be accessible in the near future.  

     In the SSM there also exist several 
other contributions to the EDMs of the neutron and the electron.  
At two-loop level, as long as the squarks and sleptons are much
heavier than the charginos and neutralinos, the diagrams 
with the squarks or sleptons are neglected compared to the 
diagrams related to the $W$-boson EDM.  
This may be indeed the case, provided that 
the CP-violating phase intrinsic in the SSM 
is of order unity, and thus  
the squarks and sleptons are heavier than 1 TeV.  
On the other hand, 
the one-loop diagrams with the charginos 
and the squarks or sleptons  
make contributions to the EDM of the neutron or  
electron, which can be larger than the two-loop contributions.  
However, 
if the squarks and sleptons are around 10 TeV, the one-loop and the 
two-loop contributions could become comparable.  
Furthermore, these one-loop and two-loop  
contributions turned out to have the same sign.  
Therefore, the neutron and 
electron EDMs arising from the two-loop diagrams give theoretical lower 
bounds for given parameter values of the gauge-Higgs sector.   

\vskip 0.5 true cm
     One of the authors (N.O.) thanks D. Schildknecht 
for helpful conversations during his visit at University of 
Bielefeld.  
We thank M. Kitahara, M. Marui, T. Saito, 
and A. Sugamoto for discussions.

\end{document}